\documentclass[11pt]{article}
\usepackage{moriond,epsfig}

\def\be{\begin{equation}}
\def\ee{\end{equation}}
\def\bea{\begin{eqnarray}}
\def\eea{\end{eqnarray}}

\begin{document}
\vspace*{4cm}
\title{The new generation CMB B-mode polarization experiment: POLARBEAR}

\author{J. Errard$^{a}$, P.A.R. Ade$^{c}$, A. Anthony$^{d}$, K. Arnold$^{b}$, F. Aubin$^{h}$, D. Boettger$^{e}$, J. Borrill$^{f,g}$, C. Cantalupo$^{f}$, M.A. Dobbs$^{h}$, D. Flanigan$^{b}$, A. Ghribi$^{b}$, N. Halverson$^{d}$, M. Hazumi$^{i}$, W.L. Holzapfel$^{b}$, J. Howard$^{b}$, P. Hyland$^{b}$, A. Jaffe$^{j}$, B. Keating$^{e}$, T. Kisner$^{f}$, Z. Kermish$^{b}$, A.T. Lee$^{b, k}$, E. Linder$^{k}$, M. Lungu$^{b}$, T. Matsumura$^{i}$, N. Miller$^{e}$, X. Meng$^{b}$, M. Myers$^{b}$, H. Nishino$^{i}$, R. OÕBrient$^{b}$, D. OÕDea$^{j}$, C. Reichardt$^{b}$, I. Schanning$^{e}$, A. Shimizu$^{i}$, C. Shimmin$^{b}$, M. Shimon$^{e}$ , H. Spieler$^{k}$, B. Steinbach$^{b}$, R. Stompor$^{a}$, A. Suzuki$^{b}$, T. Tomaru$^{i}$, H.T. Tran$^{b}$, C. Tucker$^{c}$, E. Quealy$^{b}$, P.L. Richards$^{b}$, O. Zahn$^{b, k}$, \vspace*{0,2cm}}


\address{\footnotesize $^{a}$Laboratoire Astroparticule \& Cosmologie (APC), Universit\'e Paris 7, France\\
$^{b}$Department of Physics, University of California, Berkeley U.S.\\
$^{c}$School of Physics and Astronomy, University of Cardiff, U.K.\\
$^{d}$Department of Astrophysical and Planetary Sciences, University of Colorado, U.S.\\
$^{e}$Center for Astrophysics and Space Sciences, University of California, San Diego, U.S.\\
$^{f}$Computational Cosmology Center, Lawrence Berkeley National Laboratory, U.S.\\
$^{g}$Space Sciences Laboratory, University of California, Berkeley, U.S.\\
$^{h}$Physics Department, McGill University, Canada\\
$^{i}$High Energy Accelerator Research Organization (KEK), Tsukuba, Ibaraki, Japan\\
$^{j}$Department of Physics, Imperial College, U.K.\\
$^{k}$Physics Division, Lawrence Berkeley National Laboratory, Berkeley, U.S.}

\maketitle\abstracts{We describe the Cosmic Microwave Background (CMB) polarization experiment called \textsc{Polarbear}. This experiment will use the dedicated Huan Tran Telescope equipped with a powerful 1,200-bolometer array receiver to map the CMB polarization with unprecedented accuracy. We summarize the experiment, its goals, and current status.}

\begin{figure}[htbp]
	\centering
		\includegraphics[width=8cm]{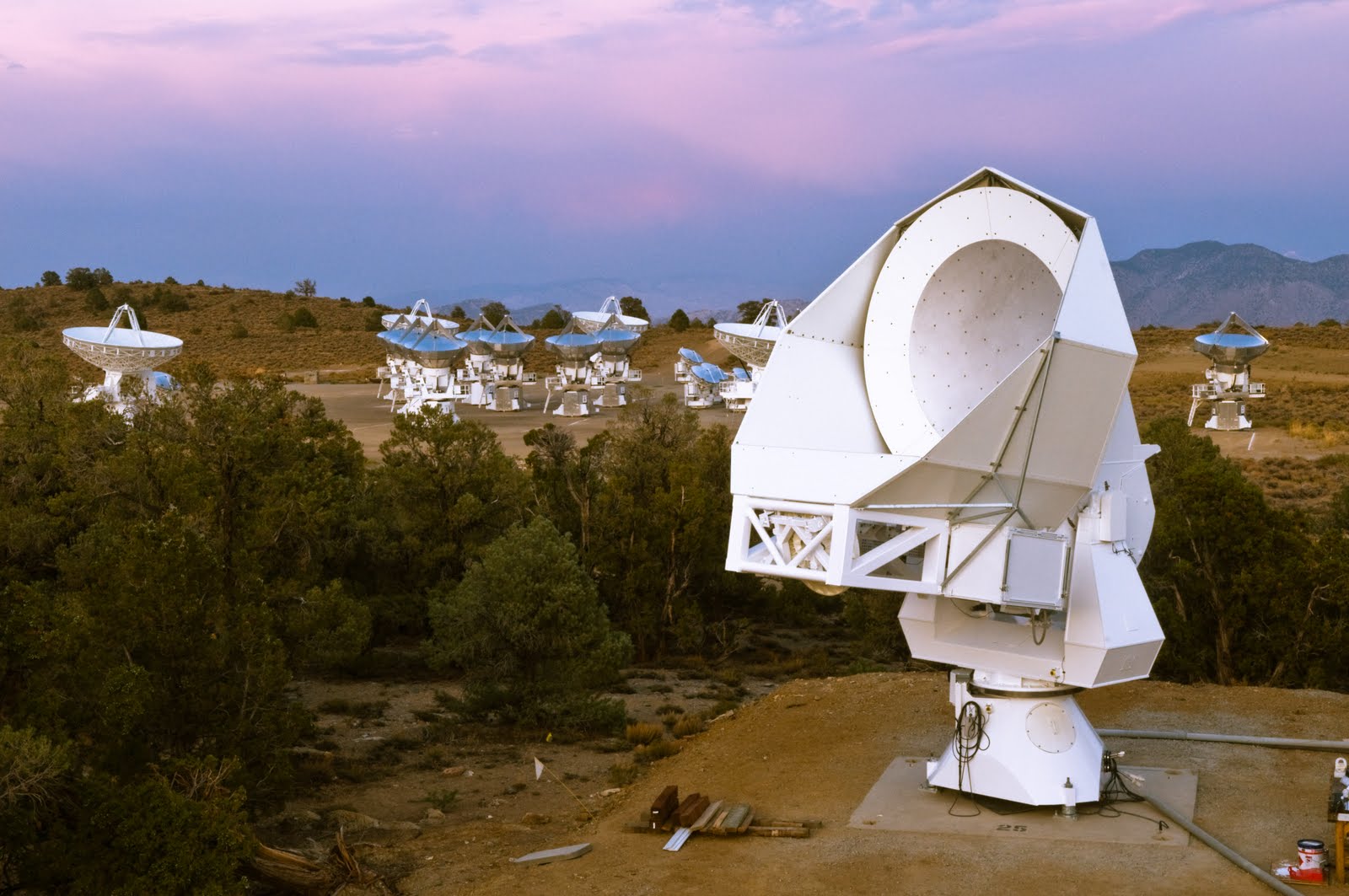}
		\caption{\footnotesize{Picture of the \textsc{Polarbear} experiment with the Huan Tran Telescope, as of February, 2010, at the Carma site, California.}}
	\label{fig:telescope}
\end{figure}


\section{\textsc{Polarbear} science goals}

Observations of the Cosmic Microwave Background (CMB) polarization anisotropies are one of the most promising and efficient tools for probing the early Universe. Especially exciting is the existence of gravitational waves that could have effected the polarization patterns imprinted on the surface of last scattering. In particular, inflation theories predict these specific perturbations. The aim of the new generation CMB experiments such as \textsc{Polarbear} is to characterize the CMB polarization and constrain model parameters. Specifically, this means detection of the E- and B-modes: the latter will provide information about structures formation (\textit{e.g.}, total neutrino mass, early action of dark energy) and about the first instant of the Universe (\textit{e.g.}, energy scale of inflation).

\section{\textsc{Polarbear} deployment}

A scaled-down version of \textsc{Polarbear} was deployed at an Eastern California site in the spring 2010 (see figure \ref{fig:telescope}). The purpose of this engineering run is to understand and model the performance of the instrument and all its subsystems to assist the development of the final, mature version of the observatory. It is also crucial in the optimization of some instrumental parameters (detectors, pointing, etc.) as well as its operations to maximize the science outcome.

After this testing phase, in the fall of 2010, \textsc{Polarbear} will be set up in the Atacama desert, Chile. It will observe 1.5 \% of the sky in 2011 and 2012.

\section{A reduced systematics-design experiment}
\label{design}

\textsc{Polarbear} is one of a new generation of CMB B-mode experiments and is designed on two pillars: \emph{sensitivity} and \emph{control of systematic instrumental effects} . On the one hand, \textsc{Polarbear} will use a high number of multifrequencies pixels. On the other hand, it will have reduced beam systematics thanks to 4' beams, a rotating Half Wave Plate to modulate the incoming polarization, low sidelobes optics, and the observed sky patches will be coordinated with the QUIET experiment. \textsc{Polarbear} design should allow to the detection of primordial B-mode with a ratio tensor-to-scalar $r\sim0.025$.\\

In its final version, \textsc{Polarbear} will observe the sky from the ground, scanning it repetitively at constant elevation with the polarization of the incident radiation modulated by a rotating half-wave plate \cite{tomo}$^{,\;}$\cite{collins} (it will rotate continuously or will be stepped once every several hours). The observation will be performed with roughly 1,200 bolometric detectors operating simultaneously (paragraph \ref{detect}) for one year. As a consequence, the \textsc{Polarbear} instrument includes a complex detector/read-out systems (paragraph \ref{multiplex}). 
Its operation will involve multiple modulations on different timescales to allow for discrimination of the sky signals from the instrumental or atmospheric ones. The instrument will observe the sky in multiple frequency bands (it will observe with the initial array at 150 GHz for a year and then switch to an array entirely at 220 GHz) to permit cleaning of the cosmological signal of interest from various astrophysical contaminations.

Below we describe some of the sub-systems of the instrument.

\subsection{Detectors Array}
\label{detect}

\begin{figure}
	\centering
		\includegraphics[width=8cm]{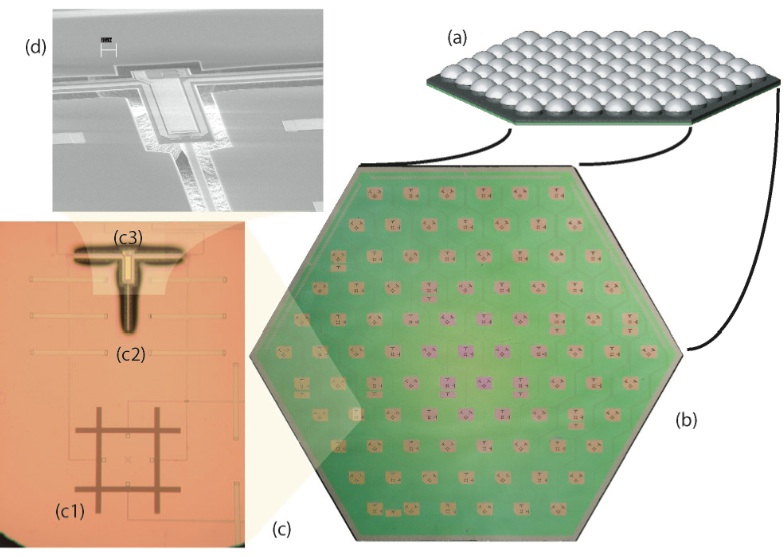}
		\caption{\footnotesize{A single focal plane sub-array, showing the array of contacting dielectric lenslets (a) each with its own anti-reflection coating. Below this is the device wafer (b), a hexagonal array of dual-polarization antennas, each coupled to 2 bolometers through microstrip filters. The device wafer is 80 mm across. (c) A picture taken under an optical microscope showing (c1) the dual polarization antenna, (c2) the microstrip filtering for one of the polarizations, and (c3) the bolometer for one of the polarizations. (d) A scanning electron microscope image of the bolometer, showing that it is mechanically and thus thermally released from the silicon substrate. The released bolometer is 80 $\mu$m across.}}
	\label{fig:detector}
\end{figure}

The Berkeley group has successfully created a crossed double slot dipole, TES Bolometer which can measure polarized radiation with high sensitivity \cite{antenna_coupled} (see fig.\ref{fig:detector}). This was an important technological success for the \textsc{Polarbear} project, and was the first step in producing the large arrays that are required for the new CMB experiments.

\begin{itemize}
	\item \textbf{\footnotesize{Antenna/Lens combination}} - A silicon hemispherical lens is placed onto the antenna. The detector chip sits directly on the lens. This antenna/lens combination has been used extensively at these frequencies and can couple efficiently to typical telescope optics.
	\item \textbf{\footnotesize{Superconducting Microstrip}} - the antenna is connected to a transmission line, which is used to bring the incoming optical power to the detector. Commonly used materials would cause high power loss at our frequencies ($\nu \sim 150-220\, GHz$), which is intolerable if we want to detect extremely weak signals. However, a superconducting microstrip is a very low loss transmission line. Niobium seems to be a convenient choice of materials, also because it has a relatively high superconducting temperature. 
	\item \textbf{\footnotesize{Band defining microstrip filters}} - an advantage of using a microstrip to connect the antenna to the bolometer is that band defining microstrip filters can be integrated into the transmission line \cite{microstrip} (see fig.\ref{fig:detector}). In a typical millimeter wave receiver, band defining filters are metal mesh off-chip optical filters. If several bands are required, several of these off-chip filters must be used. In our detectors, the filters are integrated on the chip and different pixels can easily have different frequency sensitivities. 
	\item \textbf{\footnotesize{Bolometers}} - the bolometers are composed of a \emph{terminating resistor} and a \emph{superconducting Transition Edge Sensor (TES)}, located on an 'island' isolated from the substrate. The incoming power on the superconducting microstrip is dissipated in the load resistor as heat, and the change in temperature is measured by the TES. In order to reach the sensitivity we need, the bolometer must be thermally isolated from the silicon nitride legs and the bath temperature must be below $300$ mK. This reduces the detector noise to $NEQ/U\sim36\mu K\sqrt{s} $ for the entire focal plane.
\end{itemize}

\subsection{Multiplexing}
\label{multiplex}

Large arrays of bolometric detectors require sophisticated readout schemes. To reduce thermal loading onto the coldest stages of the experiment and to reduce the complexity of instrumenting large arrays, the readout of these arrays is multiplexed. Each sensor is biased with a sinusoidal voltage at a unique frequency. The sensor signals are thus separated in frequency space and can by summed before being readout by SQUID electronics (superconducting quantum interference device). Each sensor is placed in series with a tuned filter consisting of an inductor and a capacitor with values chosen to give center frequencies from $300$ kHz to $1$ MHz.


\subsection{Structure of the telescope and cryostat}

The optical design is driven by the need to feed the large detector array. Although this is new territory for mm-wave design, many concepts can be borrowed from optical telescope design. This design is a hybrid of traditional radio-wave and visible-wave designs. The telescope is an off-axis type, used often in radio designs and chosen for its clear aperture leading to low edge diffraction \cite{offaxis}.
As explained before, \textsc{Polarbear} has been designed to measure extremely small fluctuations in polarization. Although the Antenna-Coupled TES's are naturally sensitive to polarization, an instrument capable of measuring tiny signals in the presence of large varying signals can do much better if the signal is modulated and synchronously detected. Physicists have long used such techniques with a lock-in amplifier.
As already mentioned, \textsc{Polarbear} employs a rotating Half-Wave-Plate (continuous or stepped) to modulate the incoming polarization. The signal is then demodulated using the recorded HWP position. This technique provides immunity to a number of instrumental systematic effects, such as $1/f$ noise (if the HWP rotates continuously) or scan-synchronous noise. However, there are still unwanted  systematic effects affecting the accuracy of the telescope (atmospheric contamination, instrumental polarization, etc.) which should find answers (atmospheric models, calibration, etc.).

\begin{figure}
	\centering
		\includegraphics[width=8cm]{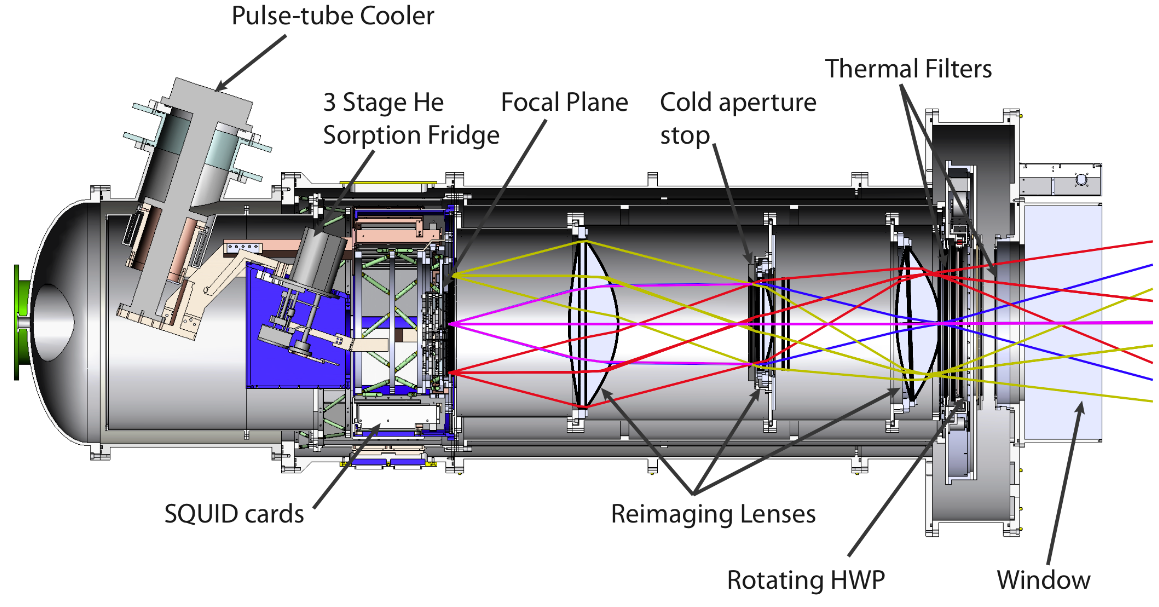}
	\caption{\footnotesize Cross section of the cryostat. We can notice the lenses, a pulse-tube cooler and a helium sorption cooler, the focal plane and the rotating Half-Wave Plate locations.}
	\label{fig:cryostat}
\end{figure}

The cryostat (fig. \ref{fig:cryostat}) is designed to cool the lenses to about 4 K and the bolometers down to 250 mK. This is accomplished with two cryogenic technologies. A pulse-tube cooler is used to bring the cold plate to about 4 K, then a helium sorption cooler brings the detector to 250 mK.

\section{Future observations and instrumentation}

 \textsc{Polarbear} will finish its engineering run in California and will be moved to the Atacama desert, in Chile, in late 2010. The focal plane will be upgraded to seven sub-arrays (about 1,200 detectors), and the experiment will begin observations in 2011. After a year of CMB observations, \textsc{Polarbear} will expand its frequency coverage to better allow examination of the data set for astrophysical foreground contamination.

\end{document}